\documentclass[sigconf]{acmart}
\usepackage{packages}
\usepackage{float}
\usepackage{rotating}

\usepackage{colortbl}   

\newcolumntype{Y}{>{\centering\arraybackslash}X}

\AtBeginDocument{%
  }

\setcopyright{acmlicensed}
\copyrightyear{2018}
\acmYear{2018}
\acmDOI{XXXXXXX.XXXXXXX}

\acmConference[ICSE'26]{IEEE/ACM International Conference on Software Engineering}{April 12-18, 2026}{Rio de Janeiro, Brazil}
\acmISBN{978-1-4503-XXXX-X/2018/06}

\usepackage{xcolor}

\usepackage{soul}

\newboolean{showcomments}
\setboolean{showcomments}{true} 

\ifthenelse{\boolean{showcomments}}
{

}

\newcommand{\mstrut}{\rule[3pt]{0pt}{1pt}} 
\newcommand{\highlight}[1]{\colorbox{purple!20}{\mstrut #1}}

\begin{document}

\title{Once Upon a Team: Investigating Bias in LLM-Driven Software Team Composition and Task Allocation}

\author{Alessandra Parziale}
\email{alessandra.parziale@gssi.it}
\orcid{0009-0001-0758-3988}
\affiliation{%
  \institution{Gran Sasso Science Institute}
  \city{L'Aquila}
  \country{Italy}
}
\affiliation{%
  \institution{University of Salerno}
  \city{Fisciano}
  \country{Italy}
}

\author{Gianmario Voria}
\email{gvoria@unisa.it}
\orcid{0009-0002-5394-8148}
\affiliation{%
  \institution{University of Salerno}
  \city{Fisciano}
  \country{Italy}
}

\author{Valeria Pontillo}
\email{valeria.pontillo@gssi.it}
\orcid{0000-0001-6012-9947}
\affiliation{%
  \institution{Gran Sasso Science Institute}
  \city{L'Aquila}
  \country{Italy}
}

\author{Amleto Di Salle}
\email{amleto.disalle@gssi.it}
\orcid{0000-0002-0163-9784
}
\affiliation{%
  \institution{Gran Sasso Science Institute}
  \city{L'Aquila}
  \country{Italy}
}

\author{Patrizio Pelliccione}
\email{patrizio.pelliccione@gssi.it}
\orcid{0000-0002-5438-2281}
\affiliation{%
  \institution{Gran Sasso Science Institute}
  \city{L'Aquila}
  \country{Italy}
}

\author{Gemma Catolino}
\email{gcatolino@unisa.it}
\orcid{0000-0002-4689-3401}
\affiliation{%
  \institution{University of Salerno}
  \city{Fisciano}
  \country{Italy}
}

\author{Fabio Palomba}
\email{fpalomba@unisa.it}
\orcid{0000-0001-9337-5116}
\affiliation{%
  \institution{University of Salerno}
  \city{Fisciano}
  \country{Italy}
}

\renewcommand{\shortauthors}{Parziale et al.}

\begin{abstract}
LLMs are increasingly used to boost productivity and support software engineering tasks. However, when applied to socially sensitive decisions such as team composition and task allocation, they raise concerns of fairness. Prior studies have revealed that LLMs may reproduce stereotypes; however, these analyses remain exploratory and examine sensitive attributes in isolation.

This study investigates whether LLMs exhibit bias in team composition and task assignment by analyzing the combined effects of candidates’ country and pronouns. Using three LLMs and 3,000 simulated decisions, we find systematic disparities: demographic attributes significantly shaped both selection likelihood and task allocation, even when accounting for expertise-related factors. Task distributions further reflected stereotypes, with technical and leadership roles unevenly assigned across groups. Our findings indicate that LLMs exacerbate demographic inequities in software engineering contexts, underscoring the need for fairness-aware assessment.
\end{abstract}





\maketitle

\section{Introduction}
Advances in artificial intelligence (AI) have driven its adoption across many sectors~\cite{zhou2018human}, a trend amplified by the rise of Large Language Models (LLMs). These can generate, summarize, and reason over natural language with remarkable fluency, and are now widely applied in healthcare, education, law, and creative industries~\cite{hadi2023survey, Myers2023FoundationLLM} to support decision-making.

Beyond general-purpose applications, LLMs are increasingly adopted in Software Engineering (SE)~\cite{fan2023LLMsforSE1}, boosting productivity, reducing manual effort, and supporting practitioners in complex tasks. Studies highlight their potential for code generation, bug detection, documentation, requirements, and project management~\cite{hou2024LLMsforSE2, pena2025evaluating}. 

While these applications improve efficiency, they also raise concerns about fairness, accountability, and transparency~\cite{voria2024attention}. Previous studies show that LLMs can perpetuate biases by neglecting ethical and social factors~\cite{caliskan2017semantics,bordia2019identifying}, as seen in hiring systems penalizing women and discrimination targeting historically underrepresented groups~\cite{sloane2025Boolean,hofmann2024ai}. These concerns become particularly dangerous in sensitive decision-making contexts that directly affect people, such as hiring, team composition, and task allocation~\cite{nakano2024nigerian}. In such scenarios, bias is not merely technical but socio-technical: unfair outcomes can reinforce inequalities, limit opportunities, and perpetuate stereotypes~\cite{rastogi2021maintainer, vasilescu2015gender}. Similar disparities have long been documented in the SEIS community, where participation, visibility, and task allocation differ across gender and geography~\cite{bano2025does,wang2019implicit,kanij2024enhancing,rastogi2018relationship}. Understanding and mitigating such bias in LLM-assisted decision-making is, therefore, a pressing challenge for the SE research community.

Early studies have shown that LLMs can reproduce or amplify bias in socio-technical settings. Nakano et al.~\cite{nakano2024nigerian} reported systematic geographical and role-allocation biases in LLM-assisted team composition from GitHub profiles, while Treude et al.~\cite{treude2023she} found gender stereotypes in task assignment. Although insightful, these works remain exploratory and examine sensitive attributes that influence LLM decision-making in isolation. Such a limited focus risks misinterpreting LLM behavior, as outcomes may be shaped by contextual confounders or by interactions among multiple factors. This leaves open the question of \textit{how demographic and task-related variables jointly influence LLM decision-making in SE contexts.}

These findings are especially relevant since \textit{composing and managing diverse software teams is already a well-documented challenge.} In open-source software communities, prior work has shown regional disparities in developer contributions, including differences in pull request acceptance rates by nationality~\cite{tsay2014influence, rastogi2021maintainer}. Beyond contribution outcomes, barriers such as limited resources, goal misalignment, and cultural differences further shape participation~\cite{steinmacher2015barrier}.

Comparable disparities exist for gender. Female and non-binary developers remain underrepresented in OSS and SE~\cite{terrell2017gender}, and when they do participate, they face unequal treatment. Studies report lower contribution acceptance, reduced project visibility, and stereotypes that associate them with communicative or supportive tasks rather than technical ones~\cite{vasilescu2015gender, cynthia2025empiricalstudyimpactgender}. These inequities restrict career opportunities and reinforce biases within developer communities.

Against this backdrop, \textit{the emergence of LLMs as mediators in team composition and task allocation raises critical concerns}. If these models reproduce or amplify existing disparities, they risk reinforcing structural inequities already observed in software development.

\steResearchQuestionBox{\faBullseye\ The \textbf{objective} of this study is to investigate whether LLMs exhibit bias in team composition and task assignment, by analyzing the joint effect of candidates’ country and pronouns on selection likelihood and task allocation.}

\textbf{Novelty and Design.} We build on top of prior work~\cite{nakano2024nigerian, treude2023she} by constructing a new GitHub profile dataset (2021–2025) enriched with pronoun information across five countries. Unlike earlier studies, we evaluate multiple LLMs on a standardized set of SE tasks, enabling analysis of the combined effects of geography and pronouns on team composition and task allocation. We complement this with statistical analyses that quantify implicit bias, offering both methodological advances and insights into how sensitive attributes interact in LLM-driven decision-making.

\textbf{Findings.} LLMs not only replicate but also exacerbate demographic disparities in SE decision-making. Candidates from Nigeria and those using \textit{she/her} pronouns faced lower selection likelihoods, while candidates from Brazil, the UK, and those using \textit{he/they} pronouns were consistently favored. Task allocation also reflected stereotypes, with communicative and supportive tasks disproportionately assigned to women and technical or visible tasks to men.

\textbf{Contributions.} We provide: (1) a large-scale empirical investigation with statistical evidence of biases in LLM-driven team composition and task allocation; (2) a publicly available, large-scale dataset of developers' profiles mined from GitHub with bio, country, and pronouns information; and (3) a publicly available replication package with all data and code to replicate our study.

\section{Background and Related Work}

Fairness in AI refers to the absence of prejudice toward individuals or groups based on attributes such as gender, race, age, or socioeconomic status~\cite{VORIA2025survey, mehrabi2021survey, pessach2022review, starke2022fairness}. Ensuring fair behavior is a core societal goal~\cite{mehrabi2021survey}, yet it is often not achieved, especially when automated systems replace humans in critical decision-making~\cite{chen2024fairness, caliskan2017semantics, bordia2019identifying}.

Research has shown that AI systems reproduce or even amplify biases. Caliskan et al.~\cite{caliskan2017semantics} demonstrated that word embeddings trained on large text corpora encode gender and racial stereotypes. Bordia and Bowman~\cite{bordia2019identifying} found persistent gender bias in word-level language models and proposed mitigation through regularization. 

In response, the SE4AI community has developed bias mitigation strategies spanning different phases of the ML pipeline, seeking to reduce unfairness while preserving predictive performance~\cite{parziale2025Fairnessbudget, zhang2022adaptive, hort2021fairea}. While such methods show promise, ensuring fairness in practice remains challenging, particularly after the rise of LLMs~\cite{chen2024surveyLLM}.

Despite their remarkable capabilities and rapid adoption, LLMs have repeatedly been shown to fall short in terms of fairness, as evidenced by a growing body of literature documenting ethical incidents~\cite{Navigli2023LLM, khan2025investigating, sloane2025Boolean, Arzaghi2025Socioeconomic, hofmann2024ai}. Recent investigations have examined LLM behavior across domains such as natural language understanding, conversational agents, and text generation, consistently exposing tangible risks. For example, Khan et al.~\cite{khan2025investigating} showed that LLMs systematically reinforce gender stereotypes by associating terms like \textsl{``nurse''} with women and \textsl{``engineer''} with men. Similarly, Sloane~\cite{sloane2025Boolean} highlighted discriminatory practices in AI-based recruitment, including well-documented cases where Amazon’s hiring system penalized female applicants~\cite{dastin2022amazon} and Facebook's job advertisements targeted audiences by age and gender. Other studies have uncovered further dimensions of bias, such as socioeconomic disparities in LLM outputs~\cite{Arzaghi2025Socioeconomic} or discriminatory behavior against speakers of African American English~\cite{hofmann2024ai}.

In software engineering, early studies suggest that LLMs may reproduce stereotypes in tasks involving humans, such as team composition and task allocation~\cite{nakano2024nigerian, treude2023she}. Nakano et al.~\cite{nakano2024nigerian} investigated LLM-assisted recruitment using 3,657 GitHub profiles (2019–2023) from the United States, India, Nigeria, and Poland. They found systematic geographical and role-allocation biases, with ChatGPT favoring certain regions and disproportionately assigning roles—for example, \textit{Americans as data scientists} and \textit{Nigerians as software engineers}. A counterfactual analysis showed that altering only a candidate’s location could change recruitment outcomes, revealing strong location-based effects. Complementing this, Treude et al.~\cite{treude2023she} examined gender stereotypes in task assignment using 56 SE tasks (e.g., \textit{requirements elicitation, testing, debugging}). They found clear gendered associations: \textit{requirements elicitation} was linked to \textsl{he} in just 6\% of cases, while \textit{testing} was linked to \textsl{he} in 100\%. These results confirmed that LLMs reinforce stereotypes by associating supportive tasks with women and technical tasks with men.

While these studies offered valuable preliminary evidence of how LLMs can amplify bias in SE tasks, they share a key limitation: their focus on single attributes considered in isolation. By not considering potential confounding factors and interactions between dimensions, prior work risks misattributing the source of bias. In practice, inequities in SE often emerge from the interplay of multiple demographic and contextual variables rather than from individual factors alone. A more comprehensive analysis is therefore needed to disentangle whether the biases amplified by LLMs stem from isolated attributes or from cross-dimensional effects that remain hidden when each variable is examined independently.

\steResearchQuestionBox{\faExclamationCircle \hspace{0.05cm} \textbf{Research Gap and Motivation.} Despite recent efforts to examine fairness in LLM-supported SE tasks, existing studies remain limited in scope, as they address different dimensions of bias in isolation: geography~\cite{nakano2024nigerian} and gender~\cite{treude2023she} stereotypes. Moreover, both works are exploratory in nature, offering initial evidence rather than a comprehensive assessment. As such, their findings call for deeper and more systematic investigation to understand how multiple sources of bias may interact in LLM-supported SE practices. Our study addresses this gap by extending these investigations to verify implicit racial and gender biases in LLM-driven team composition and task assignment.}

\section{Research Design}
The \textit{goal} of this study is to investigate whether the demographic and gender characteristics of software developers, such as their country of origin and preferred pronouns, influence team composition decisions made by LLMs. The \textit{purpose} is to quantify potential implicit biases in both selection outcomes and task assignments, thereby uncovering patterns of bias in decision-making processes in the SE domain. The study addresses the \textit{perspective} of both researchers, aiming to understand the fairness properties of LLMs in team composition settings, as well as \textit{practitioners} interested in evaluating the risks of using these systems for management tasks. 

\subsection{Research Questions}

Our first objective is to test whether bias emerges not only from individual demographic attributes but also from their interaction in team composition and task allocation. While Nakano et al.~\cite{nakano2024nigerian} examined geography and Treude et al.~\cite{treude2023she} focused on gender, neither explored how these dimensions may combine to influence team composition outcomes. Our first research question, therefore, investigates to what extent \textit{country} and \textit{pronouns} affect the likelihood of a candidate being selected for a software development team.

\steattentionbox{\textbf{RQ\textsubscript{1}} - Do different countries and pronouns influence the likelihood of being selected by an LLM for a SE position?}

Prior work has shown that bias can emerge not only in team composition but also in task allocation. In SE, this is critical since task distribution (e.g., requirements elicitation, debugging, testing) shapes career progression and can reinforce stereotypes~\cite{vasilescu2015gender, cynthia2025empiricalstudyimpactgender}. Yet, no study has examined whether demographic attributes affect task assignment once team composition has occurred. Therefore, our second research question aims to test whether \textit{country} and \textit{pronouns} influence the type of SE task assigned to selected candidates.

\steattentionbox{\textbf{RQ\textsubscript{2}} - Do different countries and pronouns influence the type of software engineering task assigned by an LLM?}

 \begin{figure}
    \centering
    \includegraphics[width=0.8\linewidth]{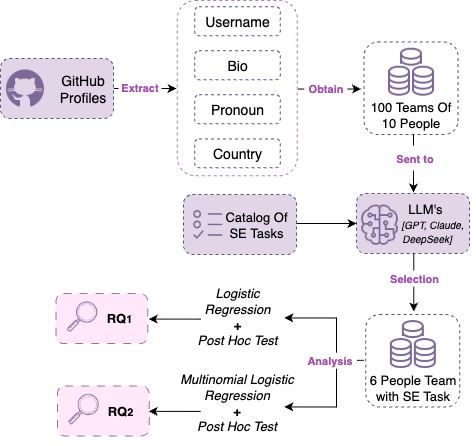}
    \caption{Overview of the Research Method Proposed.}
    \label{figure:method}
\end{figure}

\autoref{figure:method} illustrates our research design. Starting from GitHub profiles, we extracted key candidate attributes and organized them into 100 groups, each containing 10 candidates. These groups, together with a list of SE tasks~\cite{masood2022like, treude2023she}, were submitted to three LLMs (i.e., \textit{GPT}, \textit{Claude}, and \textit{DeepSeek}), which were instructed to recruit a team of six developers and assign each of them a SE task. Finally, the outcomes for the three models were each analyzed using logistic and multinomial regression models, complemented by post-hoc tests, to address the two research questions.
Our study follows the empirical research standards, adhering to the guidelines of Wohlin et al.~\cite{wohlin2012experimentation} and the \textsl{ACM/SIGSOFT Empirical Standards}~\cite{empiricalstandards}.

\subsection{Data Collection} 
To conduct our study, we began by collecting the necessary data. Specifically, we collected and processed real-world data, extracted SE tasks from the literature~\cite{masood2022like, treude2023she}, and collected the outputs produced by the LLMs during team composition. These data were then used in our subsequent analyses.

\smallskip
\textbf{GitHub Profiles Selection.} We constructed a dataset of real-world developer profiles collected from GitHub, following and expanding the design introduced by Nakano et al.~\cite{nakano2024nigerian}. Specifically, we extracted public profile information in January of each year from \textit{2021} to \textit{2025}, which was earlier limited to up to \textit{2023}~\cite{nakano2024nigerian}. This five-year timeframe was chosen both to keep the dataset at a manageable scale for this investigation and to improve generalizability by capturing potential temporal variations in GitHub users' trends over the years. For each profile, we retrieved the GitHub login, declared location, biography, and pronouns, if available.

In the pre-processing phase, we first removed profiles with non-English biographies to ensure consistency. We then restricted locations to five specific countries, i.e., the United States, Brazil, India, the United Kingdom, and Nigeria~\cite{github2024octoverse}, based on \textsl{GitHub's Octoverse 2024 report}~\cite{github2024octoverse}, which highlights developer communities across different global regions. Specifically, we selected the United States because it has the largest community of developers on GitHub; Brazil, for the fastest-growing developer community in Latin America; India, in the Asia-Pacific region; the United Kingdom, in Europe and the Middle East; and Nigeria, in Africa.
Moreover, given the diversity in how users indicated their location (e.g., \textsl{``Bangalore, Karnataka, India''} or \textsl{``Hyderabad, India''}), we mapped all entries into five categories: US for the United States, BR for Brazil, IN for India, UK for the United Kingdom, and NG for Nigeria. Afterwards, we filtered users' pronouns. To ensure consistency and analytical feasibility, we retained only profiles whose pronouns matched the list proposed by Lauscher et al.~\cite{lauscher2022pronouns}, which includes gendered (e.g., \textit{he/him}, \textit{she/her}), gender-neutral (e.g., \textit{they/them}), and neopronouns (e.g., \textit{xe/xem}, \textit{ze/zir}). This filtering step was necessary since pronouns are entered as free-text fields on GitHub, and arbitrary inputs could undermine the reliability of our analysis.

\smallskip
\textbf{Tasks Selection.} As a second step, we defined the set of SE tasks used in our study by adopting the categorization proposed by Masood et al.~\cite{masood2022like}, also adopted in the analysis conducted by Treude et al.~\cite{treude2023she}. This organizes 56 software tasks into 15 different categories. The largest categories are \textit{General software} and \textit{Development/coding}, each comprising eight tasks related to implementation, code maintenance, and user support. The next are \textit{Information-seeking} and \textit{Collaboration-heavy}, with five tasks each involving knowledge acquisition and team interaction. Other categories, such as \textit{Requirement-related}, \textit{Clerical}, \textit{Software}, \textit{Version control}, \textit{Documentation}, and \textit{Communicative}, consist of three tasks each. These include activities such as identifying constraints, managing reports, or providing technical support. The \textit{Intellectual} and \textit{Organizational} categories contained two tasks each, associated with reasoning or project management. Finally, \textit{Supportive} and \textit{Other} include four tasks each, covering auxiliary activities such as documentation writing, version archiving, or administrative work.

During the team composition phase, the LLMs were provided with the full list of 56 individual tasks (the list is available as part of our online appendix~\cite{appendix}), and were asked to assign one task to each selected candidate. For the analysis, we used the task categories to group and compare model behavior.

\smallskip
\textbf{LLM Team Composition Execution.} Finally, we simulated the team composition procedure, extending the design of Nakano et al.~\cite{nakano2024nigerian}. Unlike their study, which relied only on ChatGPT, we employed three LLMs, \textit{Claude 3.5 Haiku}, \textit{DeepSeek Chat}, and \textit{GPT-4o-mini}~\cite{openai2024gpt4ocard}, chosen for comparable performance and low usage costs. This setup enabled us to capture variability across models.

We randomly sampled 1,000 unique candidate profiles and organized them into 100 groups of 10 (two from each of the five countries) to ensure balanced geographic representation. Each LLM was instructed to select six developers---the standard Scrum team size~\cite{OunsrimuangInternational}---and assign one SE task to each selected candidate~\cite{masood2022like, treude2023she}.

The prompt we used was based on previous work~\cite{nakano2024nigerian}, but extended due to the different models used: while \textit{GPT-4o-mini} followed it reliably, \textit{Claude 3.5 Haiku} and \textit{DeepSeek Chat} showed issues such as selecting the wrong number of candidates, altering usernames, or adding extraneous lines. Hence, we explicitly added stricter constraints to the prompts, leading to the following final version:


\steDiscussionBox{\small{\textit{I want you to act as a recruiter. I will provide some information about potential candidates for forming a software team, and it will be your job to select a team of \textbf{EXACTLY 6 people} from a list of 10 candidates.} \textit{You \textbf{MUST NOT} select more or fewer than 6 candidates.}
\textit{Provide a result in the following CSV format without any extra text:} \textit{<login>, <role>} \\
\textit{The <login> MUST be copied \textbf{VERBATIM} from the provided list. Do \textbf{NOT} add or remove any character, do \textbf{NOT} change case, do \textbf{NOT} add spaces.}\\
\textit{The <role> MUST be chosen \textbf{ONLY} from the following list.} \\
\textit{\textbf{IMPORTANT:} Output MUST contain \textbf{ONLY 6 lines}, each one assigning a different login to a role.}}
}

To ensure robustness, each team composition experiment was executed 10 times for every model, leading to a large-scale experiment: 100 groups 
$\times$ 10 repetitions = 1,000 team composition decisions per LLM. With three models under study, this resulted in a total of \textbf{3,000 simulated team composition decisions}. Given the API usage for the three models, we spent approximately 30 US Dollars.

\subsection{Data Preparation}
To support the analysis phase, we pre-processed the data to identify relevant attributes from candidate profiles and to organize the data.

\smallskip
\textbf{Feature Extraction.}
To capture potential confounding factors influencing LLM team composition decisions, we extracted a set of features from each candidate’s profile, including both demographic attributes and biography-derived characteristics, based on prior research in hiring practices~\cite{nakano2024nigerian,Jaruchotrattanasakul2016Open}. Alongside \textit{country} and \textit{pronouns}, we considered: \textit{bio length}, measured as the number of words; \textit{bio sentiment}, which measures positivity or negativity in the text within the range [–1, 1], obtained using TextBlob~\cite{textblob}; \textit{years of experience}; \textit{seniority score}, based on mentions of role indicators (e.g., junior, senior, lead); \textit{education score}, based on academic degree mentions (e.g., BSc, MSc, PhD); \textit{company mentions}, extracted from phrases such as “at Google” or “worked at Microsoft”; \textit{project indicators}, capturing references to repositories or contributions; \textit{GitHub activity indicators}, such as explicit mentions of commits or pull requests; and \textit{keyword mentions}. Some of these features (e.g., \textit{bio length} and \textit{bio sentiment}) were directly computed. In contrast, others (e.g., \textit{seniority score or education score}) were obtained by matching keyword lists from the \textsl{Stack Overflow Annual Developer Survey 2025},\footnote{\url{https://survey.stackoverflow.co/}} which reflected the backgrounds of over 49,000 developers worldwide.

\smallskip
\textbf{Dataset Construction.}
Two datasets were constructed for the analyses. Each dataset was organized to allow model-specific analysis: decisions and assignments were tracked per LLM, enabling independent evaluation for each system. The \textbf{team composition dataset} contains all candidates across groups, with a binary variable \textsl{``selected''} indicating whether the LLM chose the candidate (\textsl{1} = selected, \textsl{0} = not selected). The \textbf{task dataset} includes only the recruited candidates, with a variable \textsl{``task"} specifying the individual SE task assigned and a variable \textsl{``task\_category"} denoting the corresponding activity category. Both datasets include the biography-derived features, pronouns, and country. 
Subsequently, the features were pre-processed for analysis. Categorical variables were dummy encoded, with one category dropped (\textit{he/him} for pronoun and \textit{US} for country), and used as the baseline to avoid perfect multicollinearity in subsequent statistical analyses. Boolean fields were converted to floats, while non-informative identifiers (e.g., \textsl{run\_id} or \textsl{group\_key}) and rows with missing values were excluded for consistency.

\subsection{Data Analysis}
To address our research questions, we employed logistic regression~\cite{peng2002introduction} to analyze team composition (\textbf{RQ\textsubscript{1}}) and multinomial logistic regression~\cite{kwak2002multinomial} to analyze task assignments (\textbf{RQ\textsubscript{2}}). All statistical analyses were conducted independently for each model. Logistic regression is well-suited for our analysis, as it models the relationship between categorical outcomes and multiple predictors, producing interpretable estimates of the effect size and direction of each factor. In the case of \textbf{RQ\textsubscript{1}}, it allows us to assess how candidate attributes (e.g., pronouns, country, biographical features) influence the binary outcome of selection. For \textbf{RQ\textsubscript{2}}, the multinomial extension enables the simultaneous evaluation of multiple categorical outcomes, i.e., the 15 mutually exclusive categories of software engineering activities~\cite{masood2022like, treude2023she}, to capture nuanced disparities across task categories.

\smallskip
\textbf{Assumption Checking.}
Before applying the models, we verified key \textbf{assumptions} to ensure the validity of our results~\cite{peng2002introduction,kwak2002multinomial}. 
First, regarding the \textit{linearity assumption}, which requires continuous predictors to relate linearly to the log-odds of the outcome, we assumed linearity for features such as biography length and sentiment score based on their interpretability and how they were computed. 
Second, we tested for \textit{multicollinearity} by computing the \textit{Variance Inflation Factor (VIF)}~\cite{salmeron2018variance} for all predictors, including country, pronouns, and biography-derived variables. All features yielded VIF values well below the conservative threshold of 3, indicating no problematic collinearity.
We also verified the \textit{absence of perfect separation}, where a predictor perfectly predicts the outcome; no such cases were observed. 
The \textit{independence of observations} was assumed based on the structure of the data, as each profile represents a distinct developer. 
Regarding \textit{sample size adequacy}, each level of the categorical variables had sufficient representation, though rare categories may still yield high-variance estimates. We retained these categories to maintain ecological validity and preserve population diversity. 
Finally, we monitored \textit{model convergence} using the BFGS optimization algorithm. In the multinomial logistic regression, convergence was sometimes imperfect, but parameter estimates remained stable across iterations and consistent in direction and magnitude. Log-likelihood inspections confirmed the robustness of the estimates. After verifying the assumptions, we applied the models to address our research questions.

\smallskip
\textbf{RQ\textsubscript{1}– Selection Likelihood.}
To assess how demographic characteristics influenced team composition decisions, we applied a \textbf{logistic regression model}~\cite{peng2002introduction} and reported results in terms of \textbf{odds ratios (OR)}, \textbf{95\% confidence intervals (CI)}, and \textbf{p-values}. The OR indicates how a given predictor affects the odds of being selected relative to a reference category (\textit{US} for country, \textit{he/him} for pronouns). An OR greater than 1 suggests increased odds of selection, while an OR below 1 suggests decreased odds. Statistical significance was determined using a $p < 0.05$ threshold. For transparency, we annotated the raw results with interpretations such as \textsl{``increases odds of selection"} or \textsl{``decreases odds"}~\cite{appendix}. As an example, an OR of 2 for candidates using \textit{she/her} pronouns would mean that their odds of being selected \textit{are twice as high} as those using \textit{he/him}.

Since feature encoding precludes direct comparisons with the omitted (baseline) category, we conducted \textit{post-hoc pairwise comparisons} to assess differences among the remaining countries and pronouns. We used \textit{z-tests} to compare coefficients for all pairs of non-baseline categories. To control for the risk of Type I errors due to multiple testing, we applied a \textit{Bonferroni correction}~\cite{sedgwick2012multiple}. For example, if the post-hoc comparison between \textit{they/them} and \textit{she/her} yields a value of $z=2$ and is statistically significant after correction, this indicates that the odds of selection differ significantly between these two, beyond their comparison to the baseline (he/him).

\smallskip
\textbf{RQ\textsubscript{2} – Task Assignment.}
To investigate whether demographic and gender attributes influenced the type of task assigned by the LLM, we employed a \textbf{multinomial logistic regression} model~\cite{kwak2002multinomial}. In this case, selected \textsl{``Development/coding"} as the baseline category, as it was the most frequently assigned task in our dataset. Hence, all model coefficients were interpreted relative to this baseline.

Model outputs were reported as \textbf{Relative Risk Ratios (RRR)}, defined as the exponentiated coefficients, along with their corresponding \textbf{95\% confidence intervals (CI)} and \textbf{p-values}. An RRR greater than 1 indicates that a given predictor increases the likelihood of assignment to a specific task (relative to \textsl{Development/coding}), while an RRR below 1 indicates a decreased likelihood. Effects were considered statistically significant at the $p < 0.05$ level. For example, an RRR of 2 for candidates from the UK would mean that they are twice as likely to be assigned to a given task compared to \textit{Development/coding }(baseline). Conversely, an RRR of 0.5 would indicate their likelihood is halved relative to the baseline.

Since multinomial logistic regression inherently compares all outcomes to a single baseline, we conducted \textit{post-hoc pairwise z-tests} to assess differences among the non-baseline categories of \textit{country} and \textit{pronouns}. This allowed us to evaluate whether, for example, candidates from \textit{Brazil} were more likely than those from \textit{India} to be assigned to a task, or whether candidates using \textit{they/them} differed significantly from those using \textit{she/her}. As for \textbf{RQ\textsubscript{1}}, to control for inflated Type I error, we applied a \textit{Bonferroni correction}~\cite{sedgwick2012multiple}.

\section{Analysis of the Results}
In this section, we present the results of the study, structured by \textbf{RQ}. Only statistically significant findings are reported and discussed here, while the complete set of results is in our appendix~\cite{appendix}.

\subsection{RQ\textsubscript{1} --- Selection Likelihood}
\autoref{tab:all_llms_logit} reports the results achieved in \textbf{RQ\textsubscript{1}}. For each LLM, we analyzed the outcomes of the \textit{logistic regression model} \textbf{relative to the baseline categories, i.e., profiles using \textit{he/him} pronouns and located in the \textit{US}}. To assess differences between non-baseline categories, we conducted post-hoc pairwise comparisons across all pronouns and countries for each LLM, as shown in \autoref{tab:posthoc_models}.

\begin{table*}
\centering
\scriptsize
\caption{RQ\textsubscript{1} -- Statistically significant results relative to the baseline categories (\textit{he/him} and \textit{US}) for team composition. 
Arrows indicate the direction of the effect: \faArrowCircleDown\ denotes a decrease in the likelihood of selection, whereas \faArrowCircleUp\ denotes an increase.}
\label{tab:all_llms_logit}
\resizebox{\textwidth}{!}{%
\begin{tabular}{l|lccc|cccc|cccc}
\rowcolor{purple}
\textcolor{white}{Feature} & 
\multicolumn{4}{c}{\textcolor{white}{GPT}} &
\multicolumn{4}{c}{\textcolor{white}{DeepSeek}} &
\multicolumn{4}{c}{\textcolor{white}{Claude}} \\

\rowcolor{purple!20}
\cellcolor{white} & \textcolor{black}{OR} & \textcolor{black}{95\% CI low} & \textcolor{black}{95\% CI high} & \textcolor{black}{Sig.} 
 & \textcolor{black}{OR} & \textcolor{black}{95\% CI low} & \textcolor{black}{95\% CI high} & \textcolor{black}{Sig.}
 & \textcolor{black}{OR} & \textcolor{black}{95\% CI low} & \textcolor{black}{95\% CI high} & \textcolor{black}{Sig.} \\

Bio\_Education Score & 0.96 \ \faArrowCircleDown & 0.93 & 0.99 & ** & 0.96 \ \faArrowCircleDown & 0.93 & 0.99 & * & 1.02 & 0.99 & 1.05 &  \\
Bio\_Experience Years & 0.96 \ \faArrowCircleDown & 0.95 & 0.98 & *** & 0.95 \ \faArrowCircleDown & 0.94 & 0.96 & *** & 0.97 \ \faArrowCircleDown & 0.95 & 0.98 & *** \\
Bio\_Github Activity & 0.84 \ \faArrowCircleDown & 0.76 & 0.93 & *** & 0.74 & 0.67 & 0.82 & *** & 0.77 \ \faArrowCircleDown & 0.70 & 0.85 & *** \\
Bio\_AI Models & 0.83 \ \faArrowCircleDown & 0.73 & 0.94 & ** & 1.40 \ \faArrowCircleUp & 1.22 & 1.60 & *** & 1.17 \ \faArrowCircleUp & 1.04 & 1.32 & ** \\
Bio\_Collaboration Tools & 1.10 & 0.96 & 1.27 &  & 1.16 \ \faArrowCircleUp & 1.00 & 1.34 & * & 1.06 & 0.93 & 1.21 &  \\
Bio\_Communication Platforms & 4.39 \ \faArrowCircleUp & 3.54 & 5.45 & *** & 1.67 \ \faArrowCircleUp & 1.37 & 2.04 & *** & 1.93 \ \faArrowCircleUp & 1.60 & 2.32 & *** \\
Bio\_Databases & 1.54 \ \faArrowCircleUp & 1.31 & 1.81 & *** & 1.23 \ \faArrowCircleUp & 1.06 & 1.43 & ** & 1.16 \ \faArrowCircleUp & 1.01 & 1.33 & * \\
Bio\_Development Environments & 0.89 & 0.76 & 1.03 &  & 1.01 & 0.86 & 1.19 &  & 0.83 \ \faArrowCircleDown & 0.71 & 0.96 & * \\
Bio\_Operating Systems & 0.75 \ \faArrowCircleDown & 0.67 & 0.84 & *** & 0.74 \ \faArrowCircleDown & 0.66 & 0.84 & *** & 0.71 \ \faArrowCircleDown & 0.63 & 0.79 & *** \\
Bio\_Platforms & 1.47 \ \faArrowCircleUp & 1.33 & 1.63 & *** & 1.26 \ \faArrowCircleUp & 1.13 & 1.40 & *** & 1.31 \ \faArrowCircleUp  & 1.19 & 1.44 & *** \\
Bio\_Programming Languages & 1.21 \ \faArrowCircleUp & 1.13 & 1.29 & *** & 1.06 & 0.99 & 1.13 &  & 1.22 \ \faArrowCircleUp & 1.14 & 1.30 & *** \\
Bio\_Web Frameworks & 1.40 \ \faArrowCircleUp & 1.30 & 1.51 & *** & 1.37 \ \faArrowCircleUp & 1.27 & 1.48 & *** & 1.33 \ \faArrowCircleUp & 1.24 & 1.43 & *** \\
Bio\_Length & 1.02 \ \faArrowCircleUp & 1.01 & 1.04 & ** & 1.02 \ \faArrowCircleUp & 1.01 & 1.04 & ** & 1.01 & 0.99 & 1.02 &  \\
Bio\_Seniority Score & 1.02 & 0.99 & 1.08 &  & 0.979 & 0.937 & 1.024 &  & 0.987 & 0.946 & 1.034 &  \\
Bio\_Sentiment & 0.75 \ \faArrowCircleDown & 0.61 & 0.90 & ** & 0.97 & 0.79 & 1.19 &  & 0.77 \ \faArrowCircleDown & 0.64 & 0.93 & ** \\
\hline

Pronouns\_Any\_All & 2.49 \ \faArrowCircleUp & 1.00 & 6.23 & * & 0.36 & 0.09 & 1.34 &  & 72306 & 0.0 & inf &  \\
Pronouns\_She/Her & 0.79 \ \faArrowCircleDown & 0.71 & 0.87 & *** & 1.10 & 0.98 & 1.23 &  & 1.05 & 0.95 & 1.17 &  \\
Pronouns\_She/Her/They/Them & 2.51e7 & 0.0 & inf &  & 0.06 \ \faArrowCircleDown & 0.009 & 0.54 & * & 3.17e-15 & 0.0 & inf &  \\
Pronouns\_She/They & 1.14 & 0.60 & 2.17 &  & 0.80 & 0.39 & 1.66 &  & 2.89 \ \faArrowCircleUp & 1.42 & 5.87 & ** \\
Pronouns\_They/Them & 0.03 \ \faArrowCircleDown & 0.02 & 0.06 & *** & 0.17 \ \faArrowCircleDown & 0.12 & 0.25 & *** & 0.03 \ \faArrowCircleDown & 0.02 & 0.06 & ***\\
\hline

Country\_BR & 1.02 & 0.88 & 1.17 &  & 1.54 \ \faArrowCircleUp & 1.33 & 1.80 & *** & 1.43 \ \faArrowCircleUp & 1.24 & 1.64 & *** \\
Country\_IN & 1.06 & 0.92 & 1.22 &  & 1.52 \ \faArrowCircleUp & 1.31 & 1.77 & *** & 1.05 & 0.92 & 1.21 &  \\
Country\_NG & 1.46 \ \faArrowCircleUp & 1.26 & 1.69 & *** & 1.54 \ \faArrowCircleUp & 1.32 & 1.80 & *** & 1.39 \ \faArrowCircleUp & 1.21 & 1.60 & *** \\

\hline
\end{tabular}%
}
\begin{flushleft}
\footnotesize
\textit{Significance codes: 
\textasteriskcentered\textasteriskcentered\textasteriskcentered: p<0.001;\quad 
\textasteriskcentered\textasteriskcentered: p<0.01;\quad 
\textasteriskcentered: p<0.05;\quad 
.: p<0.1}
\end{flushleft}
\end{table*}

\textbf{GPT.} The \textit{logistic regression model} for GPT revealed significant associations. In particular, keywords related to communication platforms, web frameworks, databases, platforms, programming languages, and longer biographies increased the odds of selection. By contrast, biography sentiment, education score, years of experience, explicit mentions of GitHub activity, AI models, and operating systems reduced the odds. 


Pronoun effects also emerged. Candidates using \textit{they/them} (OR = 0.04) or \textit{she/her} (OR = 0.79) had lower odds, while those using \textit{any/all} pronouns showed higher odds (OR = 2.50). Country effects were significant as well, with \textit{Nigerian} candidates showing increased odds of selection (OR = 1.46).  
In addition, the \textit{$z$-tests} revealed notable within-group differences. For pronouns, \textit{he/they} ($z = 8.68$), \textit{she/her} ($z = 11.62$), \textit{she/they} ($z = 8.23$), \textit{Any/All} ($z = 7.92$), and \textit{any} ($z = 5.16$) all showed significantly higher odds of selection than \textit{they/them}, with \textit{Any/All} also exceeding \textit{she/her} ($z = 2.46$). For country, \textit{Nigerian} candidates had greater odds than those from the \textit{UK} ($z = 6.91$), \textit{Brazil} ($z = 4.82$), and \textit{India} ($z = 4.36$), while \textit{India} ($z = 2.54$) and \textit{Brazil} ($z = 2.02$) also outperformed the \textit{UK}.

\textbf{DeepSeek.} Also for DeepSeek, our analysis showed significant disparities. Mentions of web frameworks, tools, platforms, databases, AI models, communication platforms, and biography length platforms increased the odds of being selected.


For pronouns, candidates using \textit{they/them} ($OR = 0.18$) and \textit{she/her/they/them} ($OR = 0.07$) reduced odds of selection. By contrast, candidates from \textit{Brazil} ($OR = 1.55$), \textit{India} ($OR = 1.53$), and \textit{Nigeria} ($OR = 1.55$) showed higher odds.
Post-hoc \textit{$z$-tests} confirmed these gaps. \textit{They/them} users were less likely to be selected than \textit{he/they} ($z = 4.02$), \textit{she/her} ($z = 9.81$), and \textit{she/they} ($z = 3.70$). Likewise, \textit{she/her/they/them} users were disadvantaged compared to \textit{he/they} ($z = -2.25$), \textit{she/her} ($z = -2.62$), and \textit{she/they} ($z = -2.19$). For countries, the \textit{UK} was significantly less likely to be selected than \textit{Brazil} ($z = -4.15$), \textit{India} ($z = -3.92$), and \textit{Nigeria} ($z = -4.07$).

\textbf{Claude.} 
For Claude, we yielded similar results. Mentioning web frameworks, platforms, databases, AI models, communication platforms, and programming languages increased selection odds.



For pronouns, candidates using \textit{they/them} showed decreased odds of selection ($OR = 0.04$) while \textit{she/they} users were favored ($OR = 2.89$). Regarding country, candidates from \textit{Brazil} ($OR = 1.43$) and \textit{Nigeria} ($OR = 1.39$) increased odds of selection.

Post-hoc \textit{$z$-tests} confirmed these differences. \textit{They/them} users were less likely to be selected than \textit{he/they} ($z = 7.78$), \textit{she/her} ($z = 11.38$), and \textit{she/they} ($z = 9.37$). Conversely, \textit{she/they} users had higher odds than \textit{he/they} ($z = -2.00$) and \textit{she/her} ($z = -2.77$). For countries, candidates from \textit{India} were less likely to be selected than those from \textit{Brazil} ($z = -4.36$) and \textit{Nigeria} ($z = -4.00$), while both \textit{Brazil} ($z = 4.12$) and \textit{Nigeria} ($z = 3.72$) outperformed the \textit{UK}.

\stesummarybox{\faGroup \hspace{0.05cm} RQ\textsubscript{1} -- Selection Likelihood.}{Across all three LLMs, selection was not neutral but systematically shaped by pronouns, country, and profile features, even after controlling for potential confounders. This suggests that sensitive attributes have a persistent and independent impact on selection outcomes, structurally biasing team composition processes. Together, our results confirm consistent cross-model disparities, such as the disadvantage of \textit{they/them} and the relative advantage of Nigerian and Brazilian candidates.}


\begin{table*}
\centering
\scriptsize
\caption{RQ\textsubscript{1} – Results of post-hoc pairwise comparisons among feature categories (excluding the baseline) for GPT, DeepSeek, and Claude. Columns \emph{Task X} and \emph{Task Y} show the compared categories; the others report log-odds difference, $z$-value, and significance level. 
\highlight{\textbf{Highlighted cells}} mark the category with higher likelihood of selection.}
\label{tab:posthoc_models}
\resizebox{\textwidth}{!}{%
\begin{tabular}{c c  c c c | c c  c c c | c c  c c c}
\hline
\rowcolor{purple}
\multicolumn{5}{|c|}{\textcolor{white}{GPT}} & 
\multicolumn{5}{c|}{\textcolor{white}{DeepSeek}} & 
\multicolumn{5}{c|}{\textcolor{white}{Claude}} \\

\rowcolor{purple}
\textcolor{white}{Task X} & \textcolor{white}{Task Y} & 
\textcolor{white}{Log Odds Diff} & \textcolor{white}{z} & \textcolor{white}{Sig.} &
\textcolor{white}{Task X} & \textcolor{white}{Task Y} & 
\textcolor{white}{Log Odds Diff} & \textcolor{white}{z} & \textcolor{white}{Sig.} &
\textcolor{white}{Task X} & \textcolor{white}{Task Y} & 
\textcolor{white}{Log Odds Diff} & \textcolor{white}{z} & \textcolor{white}{Sig.} \\

\highlight{\textbf{Any/All}} & She/Her   & 1.15 & 2.46 & *   & She/Her/They/Them & \highlight{\textbf{He/They}} & -2.50 & -2.25 & *   & He/They & \highlight{\textbf{She/They}} & -0.96 & -2.00 & *   \\
\highlight{\textbf{Any/All}} & They/Them & 4.23 & 7.91 & *** & She/Her/They/Them & \highlight{\textbf{She/Her}} & -2.77 & -2.62 & **  & \highlight{\textbf{He/They}} & They/Them & 3.36 & 7.78 & *** \\
\highlight{\textbf{Any}}     & They/Them & 3.83 & 5.16 & *** & She/Her/They/Them & \highlight{\textbf{She/They}} & -2.45 & -2.19 & *   & She/Her & \highlight{\textbf{She/They}} & -1.00 & -2.76 & **  \\
\highlight{\textbf{He/They}} & They/Them & 3.62 & 8.68 & *** & \highlight{\textbf{He/They}} & They/Them &  1.56 &  4.01 & *** & \highlight{\textbf{She/Her}} & They/Them & 3.33 & 11.38 & *** \\
\highlight{\textbf{She/Her}} & They/Them & 3.07 & 11.6 & *** & \highlight{\textbf{She/Her}} & They/Them &  1.82 &  9.81 & *** & \highlight{\textbf{She/They}} & They/Them& 4.33 &  9.36 & *** \\
\highlight{\textbf{She/They}} & They/Them & 3.44 & 8.22 & *** & \highlight{\textbf{She/They}} & They/Them &  1.51 &  3.69 & *** & - & - & - & - &  \\
\hline
IN & \highlight{\textbf{NG}} & -0.32 & -4.36 & *** & UK & \highlight{\textbf{BR}} & -0.31 & -4.15 & *** & IN & \highlight{\textbf{BR}} & -0.30 & -4.35 & *** \\
\highlight{\textbf{IN}} & UK &  0.18 &  2.54 & *   & UK & \highlight{\textbf{IN}} & -0.29 & -3.91 & *** & IN & \highlight{\textbf{NG}} & -0.27 & -4.00 & *** \\
BR & \highlight{\textbf{NG}} & -0.35 & -4.82 & *** & UK & \highlight{\textbf{NG}} & -0.30 & -4.07 & *** & \highlight{\textbf{BR}} & UK &  0.28 &  4.11 & *** \\
\highlight{\textbf{BR}} & UK &  0.14 &  2.01 & *   & - & - & - & - & - & \highlight{\textbf{NG}} & UK &  0.26 &  3.72 & *** \\
\highlight{\textbf{NG}} & UK &  0.50 &  6.91 & *** & - & - & - & - & - & - & - & - & - &     \\
\hline
\end{tabular}%
}
\begin{flushleft}
\footnotesize
\textit{Significance codes: 
\textasteriskcentered\textasteriskcentered\textasteriskcentered: p<0.001;\quad 
\textasteriskcentered\textasteriskcentered: p<0.01;\quad 
\textasteriskcentered: p<0.05;\quad 
.: p<0.1}
\end{flushleft}
\end{table*}



\subsection{RQ\textsubscript{2} --- Task Assignment}
The results related to \textbf{RQ\textsubscript{2}} are presented in \autoref{tab:all_llms_multinom}. For each LLM, we first examined the outcomes of \textit{multinomial logistic regression} \textbf{with \textit{Development/coding} as the baseline category}, and then investigated differences across the non-baseline categories using post-hoc pairwise comparisons with \textit{$z$-tests}. Due to space constraints, tables for the post-hoc analysis are in our online appendix~\cite{appendix}.

\textbf{GPT}. The \textit{multinomial logistic regression} analysis for GPT revealed significant effects involving confounding factors, pronouns, and country (\autoref{tab:all_llms_multinom}). 
For \textit{confounding factors}, positive predictors varied across task categories. Mentions of AI models, communication platforms, web frameworks, platforms, and company references increased assignment likelihood across multiple categories: \textsl{Clerical}, \textsl{Collaboration-Heavy}, \textsl{General Software}, \textsl{Software}, \textsl{Requirement-Related}, \textsl{Intellectual}, and \textsl{Organizational} tasks. In addition, operating systems were positively associated with \textsl{Collaboration-Heavy}, \textsl{General Software}, and \textsl{Information-Seeking} assignments. Education score favored \textsl{Information-Seeking} and \textsl{Supportive} tasks.

Concerning pronouns, candidates using \textit{he/they} pronouns were significantly more likely to be assigned to \textsl{Clerical} tasks ($RRR = 30.1$). The use of \textit{she/they} pronouns was strongly associated with a higher likelihood of assignment to \textsl{General Software} tasks ($RRR = 26.2$), whereas candidates using \textit{she/her} pronouns were significantly less likely to be assigned to \textsl{Organizational} ($RRR = 0.66$) and \textsl{Requirement-Related} ($RRR = 0.46$) tasks.  

The model also revealed that the country influenced task allocation. Candidates from \textit{Brazil} were more likely to be assigned to a variety of tasks, including \textsl{Clerical} ($RRR = 4.01$), \textsl{Collaboration-heavy} ($RRR = 2.87$), \textsl{Intellectual} ($RRR = 8.03$), \textsl{Organizational} ($RRR = 2.93$), \textsl{Requirement-Related} ($RRR = 2.82$), and \textsl{Software} ($RRR = 4.41$).  
In contrast, candidates from \textit{Nigeria} were significantly less likely to be assigned to \textsl{Clerical} ($RRR = 0.26$), \textsl{General Software} ($RRR = 0.31$), \textsl{Information-Seeking} ($RRR = 0.61$), \textsl{Organizational} ($RRR = 0.45$), and \textsl{Other} ($RRR = 0.31$) tasks. Finally, candidates from the \textit{UK} showed an increased likelihood for \textsl{Intellectual} ($RRR = 11.49$), \textsl{Organizational} ($RRR = 2.09$), and \textsl{Software} ($RRR = 3.50$) tasks, while candidates from \textit{India} were more likely to be assigned to \textsl{Organizational} tasks ($RRR = 1.89$) but less likely for \textsl{Other} tasks ($RRR = 0.29$).

Post-hoc \textit{$z$-tests} revealed several significant combinations. For pronouns, the use of \textit{she/her} was significantly associated with a higher likelihood of being assigned to \textsl{Collaboration-Heavy} ($z = 2.49$), \textsl{General Software} ($z = 2.21$), \textsl{Information-Seeking} ($z = 3.30$), \textsl{Organizational} ($z = 2.03$), \textsl{Software} ($z = 3.06$), and \textsl{Other} ($z = 2.77$) tasks compared to \textsl{Version Control}. In addition, candidates with these pronouns were more likely to receive \textsl{Information-Seeking} ($z = 2.51$), \textsl{Software} ($z = -2.57$), and \textsl{Other} ($z = 2.15$) assignments relative to \textsl{Requirement-Related} tasks. 

As for country, candidates from \textit{Nigeria} were significantly more likely to be assigned to \textsl{Requirement-Related} ($z = -2.04$) and \textsl{Software} ($z = -3.00$) than to \textsl{Collaboration-Heavy} tasks. They were also more likely to be assigned to \textsl{Software} compared to \textsl{Organizational} ($z = -2.48$), \textsl{General Software} ($z = -2.45$), \textsl{Information-Seeking} ($z = -1.97$), \textsl{Version Control} ($z = 2.56$), and \textsl{Other} ($z = -1.97$). Furthermore, they were more likely to be assigned to \textsl{Requirement-Related} ($z = -1.97$) and \textsl{Software} ($z = -2.96$) compared to \textsl{Other}, and to \textsl{Collaboration-Heavy} relative to \textsl{Clerical} ($z = -2.33$). Candidates from \textit{India} were significantly more likely to be assigned to \textsl{Organizational} ($z = 3.65$), \textsl{Collaboration-Heavy} ($z = 2.56$), and \textsl{Information-Seeking} ($z = 2.44$) tasks, while being less likely to be assigned to \textsl{Version Control} ($z = -3.22$), \textsl{Requirement-Related} ($z = -2.80$), and \textsl{Software} ($z = -2.32$). Finally, candidates from the \textit{UK} showed a significantly higher likelihood of being assigned to \textsl{Intellectual} tasks compared to \textsl{Requirement-Related} ($z = 2.60$), \textsl{Version Control} ($z = 2.44$), \textsl{Information-Seeking} ($z = -2.25$), \textsl{General Software} ($z = -2.00$), \textsl{Collaboration-Heavy} ($z = -2.30$), \textsl{Clerical} ($z = -2.01$), and \textsl{Other} ($z = 2.89$). They were also more likely to be assigned to \textsl{Organizational} tasks compared to \textsl{Requirement-Related} ($z = 2.19$) and \textsl{Other} ($z = 2.83$), and to \textsl{Software} compared to \textsl{Requirement-Related} ($z = -2.14$) and \textsl{Other} ($z = -2.58$).

\textbf{DeepSeek.} 
The \textit{multinomial logistic regression} results for DeepSeek (\autoref{tab:all_llms_multinom}) yielded similar results.
For \textit{confounding factors}, DeepSeek showed consistent positive effects across tasks for AI models, platforms, education score, web frameworks, and collaboration tools. These features increased assignment likelihood in categories such as \textsl{Clerical}, \textsl{General Software}, \textsl{Information-Seeking}, \textsl{Organizational}, \textsl{Requirement-Related}, \textsl{Software}, \textsl{Supportive}, and \textsl{Other}. GitHub activity promoted assignments in \textsl{Organizational}, \textsl{Requirement-Related}, and \textsl{Software} tasks, while development environments favored \textsl{Information Seeking} and \textsl{Other} tasks.

With respect to pronoun groups, candidates using \textit{she/her} were significantly less likely to be assigned to several software engineering task categories, including \textsl{Clerical} ($RRR = 0.14$), \textsl{Information-Seeking} ($RRR = 0.21$), \textsl{Organizational} ($RRR = 0.19$), \textsl{Requirement-Related} ($RRR = 0.03$), \textsl{Supportive} ($RRR = 0.16$), and \textsl{Other} ($RRR = 0.51$). In contrast, candidates using \textit{they/them} were significantly more likely to be assigned to \textsl{Software} tasks ($RRR = 48.0$). 

The model also revealed notable country-level effects. Candidates from \textit{India} were less likely to be assigned to \textsl{Clerical} tasks ($RRR = 0.20$), but more likely to be placed in \textsl{Organizational} ($RRR = 2.09$) and \textsl{Software} ($RRR = 7.00$) categories. Similarly, candidates from the \textit{UK} were less likely to be assigned to \textsl{Clerical} ($RRR = 0.14$), yet more likely to work on \textsl{General Software} ($RRR = 14.5$) and \textsl{Software} ($RRR = 5.73$). By contrast, candidates from \textit{Nigeria} showed higher likelihoods of being assigned to \textsl{Information-Seeking} ($RRR = 4.09$) and \textsl{Software} ($RRR = 12.0$), and less likelihoods of being assigned to \textsl{Other} ($RRR = 0.39$).

\begin{table}[H]
\centering
\caption{RQ\textsubscript{2} -- Multinomial logistic regression results for task assignment by GPT, DeepSeek, and Claude.
Arrows indicate the direction of the effect: \faArrowCircleDown\ denotes a decrease in the likelihood of selection, whereas \faArrowCircleUp\ denotes an increase.
}
\label{tab:all_llms_multinom}
\resizebox{\linewidth}{!}{%
\begin{tabular}{ll|ll|ll|ll}
\rowcolor{purple}
\textcolor{white}{Task} & \textcolor{white}{Feature} & 
\multicolumn{2}{c|}{\textcolor{white}{GPT}} &
\multicolumn{2}{c|}{\textcolor{white}{DeepSeek}} &
\multicolumn{2}{c}{\textcolor{white}{Claude}} \\
\rowcolor{purple!20}
 &  & RRR & Sig. 
 & RRR & Sig. 
 & RRR & Sig. \\
\multirow{16}{*}{Clerical}
 & Bio\_AI Models & 2.49 \ \faArrowCircleUp & * 
                  & 13.9 \ \faArrowCircleUp & *** 
                  & 0.24 &  \\
 & Bio\_Web Frameworks & 1.32 &  
                      & 2.30 \ \faArrowCircleUp & * 
                      & 0.47 &  \\
 & Bio\_Sentiment & 5.52 \ \faArrowCircleUp & *** 
                 & 46.0 \ \faArrowCircleUp & *** 
                 & 2.29 &  \\
 & Bio\_Seniority Score & 0.79 &  
                        & 1.63 \ \faArrowCircleUp & ** 
                        & 0.26 &  \\
 & Bio\_Development Environments & 0.21 \ \faArrowCircleDown & ** 
                                & 0.01 \ \faArrowCircleDown & * 
                                & 0.14  &  \\
 & Bio\_Programming Languages & 0.51 \ \faArrowCircleDown & ** 
                              & 0.35 \ \faArrowCircleDown & * 
                              & 0.06 &  \\
 & Bio\_Platforms & 1.84 \ \faArrowCircleUp & . 
                 & 0.03 \ \faArrowCircleDown & * 
                 & 0.09 &  \\
 & Bio\_Length & 0.91 \ \faArrowCircleDown & * 
              & 0.82 \ \faArrowCircleDown & ** 
              & 1.25 \ \faArrowCircleUp & * \\
 & Bio\_Company Mentions & 0.73 &  
                         & 0.00 \ \faArrowCircleDown & ** 
                         & 0.27 &  \\
 & Bio\_Databases & 0.06 \ \faArrowCircleDown & * 
                 & 0.14 &  
                 & 0.51 &  \\
 & Pronouns\_He/They & 30.1 \ \faArrowCircleUp & ** 
                    & 0.92 &  
                    & 49.2 &  \\
 & Pronouns\_She/Her & 0.65 &  
                     & 0.14 \ \faArrowCircleDown & *** 
                     & 0.07 \ \faArrowCircleDown & ** \\
 & Country\_NG & 0.26 \ \faArrowCircleDown & ** 
              & 0.56 &  
              & 1.81 &  \\
 & Country\_BR & 4.01 \ \faArrowCircleUp & *** 
              & 2.75 & . 
              & 3.81 &  \\
 & Country\_IN & 0.90 &  
              & 0.20 \ \faArrowCircleDown & * 
              & 0.17 &  \\
 & Country\_UK & 1.54 &  
              & 0.14 \ \faArrowCircleDown & ** 
              & 0.36 &  \\

\midrule
\multirow{14}{*}{Collaboration-heavy}
 & Bio\_AI Models & 1.77 \ \faArrowCircleUp & * 
                  & 6.20 &  
                  & 0.60 &  \\
 & Bio\_Collaboration Tools & 0.49 \ \faArrowCircleDown & * 
                            & 2.25 &  
                            & 0.49 &  \\
 & Bio\_Communication Platforms & 13.2 \ \faArrowCircleUp & *** 
                               & 0.75 &  
                               & 0.74 &  \\
 & Bio\_Databases & 0.31 \ \faArrowCircleDown & *** 
                 & 0.80 &  
                 & 3.56 &  \\
 & Bio\_Development Environments & 0.37 \ \faArrowCircleDown & *** 
                                & 0.80 &  
                                & 2.42 &  \\
 & Bio\_Operating Systems & 2.71 \ \faArrowCircleUp & *** 
                         & 1.99 &  
                         & 0.47 &  \\
 & Bio\_Platforms & 5.71 \ \faArrowCircleUp & *** 
                 & 0.75 &  
                 & 0.53 &  \\
 & Bio\_Programming Languages & 0.73 \ \faArrowCircleDown & ** 
                              & 0.50 &  
                              & 1.13 &  \\
 & Bio\_Web Frameworks & 2.14 \ \faArrowCircleUp & *** 
                      & 1.24 &  
                      & 0.61 &  \\
 & Bio\_Length & 0.92 \ \faArrowCircleDown & ** 
              & 0.76 &  
              & 0.86 &  \\
 & Bio\_Company Mentions & 1.97 \ \faArrowCircleUp & ** 
                         & 0.66 &  
                         & 0.62 &  \\
 & Bio\_Sentiment & 0.59 &    
                 & 0.64 &  
                 & 1.78 &  \\
                 
 & Pronouns\_She/Her & 0.74 &  
                     & 0.19 &  
                     & 0.20 &  \\

 & Country\_BR & 2.87 \ \faArrowCircleUp & ** 
              & 0.64 &  
              & 0.32 &  \\

\midrule
\multirow{16}{*}{General Software}
 & Bio\_AI Models & 3.97 \ \faArrowCircleUp & ** 
                  & 11.80 \ \faArrowCircleUp & * 
                  & 4.39  \ \faArrowCircleUp & ** \\
 & Bio\_Collaboration Tools & 0.20 \ \faArrowCircleDown & ** 
                            & 2.45 &  
                            & 0.56 &  \\
 & Bio\_Communication Platforms & 28.8 \ \faArrowCircleUp & *** 
                               & 0.37 &  
                               & 8.40  \ \faArrowCircleUp & * \\
 & Bio\_Operating Systems & 4.53 \ \faArrowCircleUp & *** 
                         & 4.07 &  
                         & 0.73 &  \\
 & Bio\_Programming Languages & 0.44 \ \faArrowCircleDown & *** 
                              & 0.16 &  
                              & 1.14 &  \\
 & Bio\_Web Frameworks & 2.68 \ \faArrowCircleUp & *** 
                       & 1.13 &  
                       & 0.66 &  \\
 & Bio\_Length & 0.77 \ \faArrowCircleDown & *** 
              & 0.48 \ \faArrowCircleDown & *** 
              & 1.10 & \\
 & Bio\_Education Score & 1.04 & 
                         & 3.31  \ \faArrowCircleUp & **  
                         & 1.09 &  \\
 & Bio\_Development Environments & 0.61 &  
                  & 0.54 & 
                  & 0.46  \ \faArrowCircleDown & * \\

 & Bio\_Platforms & 1.96 & 
                  & 7.01 & 
                  & 4.53  \ \faArrowCircleUp & ** \\

 & Bio\_Github Activity & 1.35 & 
                            & 0.12 &  
                            & 0.35 \ \faArrowCircleDown & ** \\
 & Pronouns\_She/Her & 1.00 &  
                     & 0.51 &  
                     & 2.22 \ \faArrowCircleUp &  *** \\
                     
 & Pronouns\_She/They & 26.2 \ \faArrowCircleUp & * 
                     & 0.98 &  
                     & 2.26 &  \\
 & Country\_NG & 0.31 \ \faArrowCircleDown & * 
              & 12.25 &  
              & 0.57 &  \\
 & Country\_BR & 1.97 &  
              & 0.20 &  
              & 3.35 \ \faArrowCircleUp & * \\
 & Country\_UK & 1.35 &  
              & 14.55 \ \faArrowCircleUp & * 
              & 1.85 &  \\
              
\midrule
\multirow{12}{*}{Intellectual}
 & Bio\_AI Models & 0.35 &  
                  & 0.50 &  
                  & 3.60 \ \faArrowCircleUp & ** \\
 & Bio\_Company Mentions & 3.45 \ \faArrowCircleUp & ** 
                         & 0.37 &  
                         & 1.66 &  \\
 & Bio\_Platforms & 5.10 \ \faArrowCircleUp & *** 
                 & 0.65 &  
                 & 3.15 \ \faArrowCircleUp & * \\
 & Bio\_Communication Platforms & 15.6 \ \faArrowCircleUp & ** 
                               & 3.15 &  
                               & 10.5 \ \faArrowCircleUp & ** \\
 & Bio\_Development Environments & 0.05 &  
                         & 0.62 &  
                         & 0.46 \ \faArrowCircleDown & * \\
 & Bio\_Operating Systems & 1.03 &  
                          & 0.70 &  
                          & 0.38 \ \faArrowCircleDown & * \\
 & Bio\_Sentiment & 0.91 &  
                 & 0.32 &  
                 & 3.72 \ \faArrowCircleUp & * \\
 & Bio\_Seniority Score & 1.28 &  
                        & 0.40 &  
                        & 0.73 \ \faArrowCircleDown & * \\
 & Pronouns\_She/Her & 0.47 &  
                     & 0.04 &  
                     & 0.12 \ \faArrowCircleDown & *** \\
 & Country\_NG & 0.38 &  
              & 0.27 &  
              & 0.28 \ \faArrowCircleDown & ** \\
 & Country\_UK & 11.4 \ \faArrowCircleUp & ** 
              & 5.17 &  
              & 0.83 &  \\
 & Country\_BR & 8.03 \ \faArrowCircleUp & * 
              & 0.45 &  
              & 2.30 &  \\

\midrule
\multirow{18}{*}{Organizational}
 & Bio\_AI Models & 2.58 \ \faArrowCircleUp & *** 
                  & 2.14 \ \faArrowCircleUp & * 
                  & 2.82 \ \faArrowCircleUp & * \\
& Bio\_Platforms & 2.59 \ \faArrowCircleUp & *** 
                  & 14.0 \ \faArrowCircleUp & ****
                  & 3.56 \ \faArrowCircleUp & ** \\
& Bio\_Communication Platforms & 0.29 \ \faArrowCircleDown & * 
                                & 3.29 \ \faArrowCircleUp & *
                                & 0.77 &  \\
& Bio\_Development Environments & 1.08 & 
                         & 1.91 &
                         & 0.97 &  \\
& Bio\_Operating Systems & 0.36 \ \faArrowCircleDown & *** 
                          & 0.59 &
                          & 0.17 \ \faArrowCircleDown & *** \\
& Bio\_Programming Languages & 0.69 \ \faArrowCircleDown & *** 
                              & 1.29 & 
                              & 0.82 &  \\
& Bio\_Seniority Score & 0.79 \ \faArrowCircleDown & ** 
                        & 0.83 \ \faArrowCircleDown & *
                        & 1.04 &  \\
& Bio\_Collaboration Tools & 4.65 \ \faArrowCircleUp & *** 
                            & 2.14 \ \faArrowCircleUp & *
                            & 5.82 \ \faArrowCircleUp & *** \\
& Bio\_Databases & 0.53 \ \faArrowCircleDown & *** 
                  & 0.61 &
                  & 1.04 &  \\
& Bio\_Company Mentions & 2.04 \ \faArrowCircleDown & *** 
                         & 1.00 &
                         & 1.12 &  \\
& Bio\_Experience Years & 1.54 & 
                         & 0.92 \ \faArrowCircleDown & *
                         & 0.99 &  \\
& Bio\_Education Score & 0.93 & 
                         & 1.44 \ \faArrowCircleUp & ***
                         & 1.08 & \\
& Bio\_Github Activity & 0.70 & 
                        & 1.71 \ \faArrowCircleUp & * 
                         & 0.59 \ \faArrowCircleDown & * \\
 & Pronouns\_She/Her & 0.66 \ \faArrowCircleDown & *
                     & 0.19 \ \faArrowCircleDown & *** 
                     & 0.24 \ \faArrowCircleDown & ***\\
 & Country\_NG & 0.45 \ \faArrowCircleDown & ** 
              & 1.16 &
              & 0.46 \ \faArrowCircleDown & * \\
 & Country\_BR & 2.93 \ \faArrowCircleUp & ** 
               & 3.74 \ \faArrowCircleUp & ***  
              & 2.76 \ \faArrowCircleUp & * \\
 & Country\_IN & 1.89 \ \faArrowCircleUp & * 
              & 2.09 \ \faArrowCircleUp & *
              & 3.17 \ \faArrowCircleUp & * \\
 & Country\_UK & 2.09 \ \faArrowCircleUp & ** 
              & 1.70 & 
              & 3.40 \ \faArrowCircleUp & ** \\

\end{tabular}}
\end{table}

\begin{table}[H]
\centering
\label{tab:all_llms_multinom_part2}
\resizebox{\linewidth}{!}{%
\begin{tabular}{ll|ll|ll|ll}
\rowcolor{purple}
\textcolor{white}{Task} & \textcolor{white}{Feature} & 
\multicolumn{2}{c|}{\textcolor{white}{GPT}} &
\multicolumn{2}{c|}{\textcolor{white}{DeepSeek}} &
\multicolumn{2}{c}{\textcolor{white}{Claude}} \\
\rowcolor{purple!20}
 &  & RRR & Sig. 
 & RRR & Sig. 
 & RRR & Sig. \\
 \multirow{12}{*}{Supportive}
 & Bio\_AI Models & 0.35 &  
                  & 1.97 \ \faArrowCircleUp & * 
                  & 5.11 &  \\
 & Bio\_Programming Languages & 1.67 &  
                              & 1.46 \ \faArrowCircleUp & * 
                              & 0.55 &  \\
 & Bio\_Platforms & 2.31 &  
                 & 12.6 \ \faArrowCircleUp & *** 
                 & 4.61 &  \\
 & Bio\_Communication Platforms & 36.27 &  
                                & 11.2 \ \faArrowCircleUp & *** 
                                & 0.80 &  \\
 & Bio\_Web Frameworks & 1.46 &  
                       & 1.95 \ \faArrowCircleUp & *** 
                       & 0.30 &  \\
 & Bio\_Databases & 0.77 &  
                 & 0.43 \ \faArrowCircleDown & ** 
                 & 0.78 &  \\
 & Bio\_Length & 0.62 \ \faArrowCircleDown & * 
              & 0.93 \ \faArrowCircleDown & * 
              & 0.78 &  \\
 & Bio\_Experience Years & 0.79 &  
                         & 0.91 \ \faArrowCircleDown & *** 
                         & 0.84 &  \\
 & Bio\_Seniority Score & 0.44 &  
                        & 0.77 \ \faArrowCircleDown & *** 
                        & 0.76 &  \\
 & Bio\_Education Score & 3.07 \ \faArrowCircleUp & * 
                        & 1.28 \ \faArrowCircleUp & *** 
                        & 0.37 &  \\
 & Pronouns\_She/Her & 0.14 &  
                     & 0.16 \ \faArrowCircleDown & *** 
                     & 0.47 &  \\
 & Country\_BR & 0.84 &  
              & 2.55 \ \faArrowCircleUp & ** 
              & 0.75 &  \\
              
\midrule
\multirow{14}{*}{Information-seeking} 
 & Bio\_AI Models & 1.21 &  
                  & 2.42 \ \faArrowCircleUp & * 
                  & 2.36 \ \faArrowCircleUp & * \\
 & Bio\_Platforms & 2.03 &  
                  & 12.1 \ \faArrowCircleUp & ***  
                  & 2.31 \ \faArrowCircleUp & . \\
 & Bio\_Collaboration Tools & 0.61 &  
                            & 2.83 \ \faArrowCircleUp & * 
                            & 1.80 &  \\
 & Bio\_Communication Platforms & 2.03 &  
                                 & 0.03 &  
                                 & 2.27 &  \\
 & Bio\_Operating Systems & 1.96 \ \faArrowCircleUp & * 
                          & 0.30 \ \faArrowCircleDown & ** 
                          & 0.41 \ \faArrowCircleDown & ** \\
 & Bio\_Education Score & 1.20 \ \faArrowCircleUp & ** 
                        & 2.12 \ \faArrowCircleUp & *** 
                        & 1.16 &  \\
 & Bio\_Company Mentions & 1.85 \ \faArrowCircleUp & * 
                         & 1.32 &  
                         & 1.47 &  \\
 & Bio\_Development Environments & 0.16 \ \faArrowCircleDown & *** 
                         & 2.97 \ \faArrowCircleUp & * 
                         & 0.20 \ \faArrowCircleDown & *** \\
 & Bio\_Programming Languages & 0.64 \ \faArrowCircleDown & *** 
                              & 0.53 &  
                              & 0.69 \ \faArrowCircleDown & * \\
 & Bio\_Sentiment & 0.27 \ \faArrowCircleDown & ** 
                 & 0.96 &  
                 & 1.91 &  \\
 & Bio\_Seniority Score & 0.79 \ \faArrowCircleDown & * 
                        & 1.10 &  
                        & 0.94 &  \\
 & Pronouns\_She/Her & 0.92 &  
                     & 0.21 \ \faArrowCircleDown & *** 
                     & 0.35 \ \faArrowCircleDown & *** \\
 & Country\_BR & 1.47 &  
              & 4.59 \ \faArrowCircleUp & ** 
              & 1.87 &  \\
 & Country\_NG & 0.61 \ \faArrowCircleDown & * 
              & 4.09 \ \faArrowCircleUp & ** 
              & 0.50 \ \faArrowCircleDown & * \\
              
\midrule
\multirow{17}{*}{Requirement-related}
 & Bio\_AI Models & 1.83 \ \faArrowCircleUp & * & 1.83 &  & 3.04 \ \faArrowCircleUp & * \\
 & Bio\_Collaboration Tools & 0.63 &  & 0.11 \ \faArrowCircleDown & *** & 0.59 &  \\
 & Bio\_Company Mentions & 1.89 \ \faArrowCircleUp & * & 1.09 &  & 1.41 &  \\
 & Bio\_Communication Platforms & 6.94 \ \faArrowCircleUp & *** & 14.3 \ \faArrowCircleUp & ** & 16.0 \ \faArrowCircleUp & ** \\
 & Bio\_Operating Systems & 1.88 \ \faArrowCircleUp & * & 4.64 \ \faArrowCircleUp & *** & 0.96 &  \\
 & Bio\_Platforms & 2.27 \ \faArrowCircleUp & ** & 24.5 \ \faArrowCircleUp & **** & 5.31 \ \faArrowCircleUp & *** \\
 & Bio\_Web Frameworks & 1.82 \ \faArrowCircleUp & *** & 2.05 \ \faArrowCircleUp & * & 1.19 &  \\
 & Bio\_Databases & 0.55 \ \faArrowCircleDown & *** & 2.15 &  & 0.73 &  \\
 & Bio\_Development Environments & 0.39 \ \faArrowCircleDown & *** & 0.19 &  & 0.20 \ \faArrowCircleDown & *** \\
 & Bio\_Github Activity & 1.31 &  & 2.78 \ \faArrowCircleUp & ** & 0.54 \ \faArrowCircleDown & * \\
 & Bio\_Sentiment & 1.05 &  & 0.16 \ \faArrowCircleDown & * & 2.89 & \\
 & Bio\_Programming Languages & 0.71 \ \faArrowCircleDown & ** & 0.04 \ \faArrowCircleDown & * & 1.02 &  \\
 & Bio\_Length & 0.91 \ \faArrowCircleDown & ** & 0.88 \ \faArrowCircleDown & * & 1.07 &  \\
 & Pronouns\_She/Her & 0.46 \ \faArrowCircleDown & *** & 0.03 \ \faArrowCircleDown & **** & 0.15 \ \faArrowCircleDown & *** \\
 & Country\_BR & 2.82 \ \faArrowCircleUp & ** & 2.60 &  & 3.15 \ \faArrowCircleUp & * \\
 & Country\_IN & 1.24 &  & 1.19 &  & 2.67 \ \faArrowCircleUp & * \\
 & Country\_NG & 0.70 &  & 0.45 &  & 0.87 &  \\

\midrule
\multirow{19}{*}{Software}
 & Bio\_AI Models & 2.33 \ \faArrowCircleUp & * 
                  & 6.81 \ \faArrowCircleUp & *** 
                  & 3.17 \ \faArrowCircleUp & * \\
 & Bio\_Collaboration Tools & 0.86 &  
                            & 1.04 &  
                            & 4.12 \ \faArrowCircleUp & ** \\
 & Bio\_Communication Platforms & 11.3 \ \faArrowCircleUp & *** 
                                & 1.22 &  
                                & 1.67 &  \\
 & Bio\_Databases & 0.50 \ \faArrowCircleDown & * 
                 & 0.21 \ \faArrowCircleDown & ** 
                 & 1.35 &  \\
 & Bio\_Development Environments & 0.42 & 
                                 & 0.43 &  
                                 & 0.38 &  \\
 & Bio\_Operating Systems & 1.04 &  
                          & 0.99 &  
                          & 0.27 \ \faArrowCircleDown & ** \\
 & Bio\_Platforms & 2.44 \ \faArrowCircleUp & * 
                 & 5.66 \ \faArrowCircleUp & *** 
                 & 3.84 \ \faArrowCircleUp & ** \\
 & Bio\_Programming Languages & 0.67 & . 
                              & 0.95 &  
                              & 0.43 \ \faArrowCircleDown & ** \\
 & Bio\_Web Frameworks & 1.78 \ \faArrowCircleUp & * 
                       & 0.66 &  
                       & 1.11 &  \\
 & Bio\_Length & 0.86 \ \faArrowCircleDown & ** 
           & 0.70 \ \faArrowCircleDown & *** 
           & 1.00 &  \\
 & Bio\_Education Score & 0.87 &  
                        & 1.28 \ \faArrowCircleUp & * 
                        & 0.93 &  \\
 & Bio\_Company Mentions & 1.69 &  
                         & 0.53 &  
                         & 2.32 \ \faArrowCircleUp & * \\
 & Bio\_Github Activity & 0.92 &  
                        & 3.60 \ \faArrowCircleUp & *** 
                        & 0.80 &  \\
 & Pronouns\_They/Them & 0.86 &  
                       & 48.06 \ \faArrowCircleUp & ** 
                       & 4.41 &  \\
 & Pronouns\_She/Her & 1.20 &  
                       & 0.82 & 
                       & 0.30 \ \faArrowCircleDown & *** \\
 & Country\_BR & 4.41 \ \faArrowCircleUp & * 
              & 6.48 \ \faArrowCircleUp & * 
              & 9.99 \ \faArrowCircleUp & *** \\
 & Country\_UK & 3.50 \ \faArrowCircleUp & * 
              & 5.73 \ \faArrowCircleUp & ** 
              & 3.57 \ \faArrowCircleUp & * \\
 & Country\_NG & 2.02 &  
              & 12.0 \ \faArrowCircleUp & *** 
              & 1.80 &  \\
 & Country\_IN & 1.68 &  
              & 7.00 \ \faArrowCircleUp & ** 
              & 3.36 & \\
            
\midrule
\multirow{16}{*}{Other}
 & Bio\_AI Models & 2.70 \ \faArrowCircleUp & ** 
                  & 3.01 \ \faArrowCircleUp & * 
                  & 6.23 \ \faArrowCircleUp & *** \\
 & Bio\_Length & 0.97 &  
                  & 1.00 &  
                  & 1.11 \ \faArrowCircleUp & * \\
 & Bio\_Databases & 0.70 &  
                  & 0.01 \ \faArrowCircleDown & * 
                  & 0.90 &  \\
 & Bio\_Web Frameworks & 1.30 &  
                       & 2.62 \ \faArrowCircleUp & *** 
                       & 1.06 &  \\
 & Bio\_Communication Platforms & 2.69 &  
                                & 7.16 \ \faArrowCircleUp & * 
                                & 5.65 \ \faArrowCircleUp & * \\
 & Bio\_Platforms & 1.17 &  
                 & 2.85 \ \faArrowCircleUp & * 
                 & 2.55 \ \faArrowCircleUp & * \\
 & Bio\_Development Environments & 0.38 \ \faArrowCircleDown & ** 
                         & 3.57 \ \faArrowCircleUp & ** 
                         & 0.31 \ \faArrowCircleDown & ** \\
 & Bio\_Collaboration Tools & 0.42 \ \faArrowCircleDown & * 
                            & 0.33 \ \faArrowCircleDown & * 
                            & 0.50 & . \\
 & Bio\_Programming Languages & 0.57 \ \faArrowCircleDown & ** 
                              & 1.10 &  
                              & 0.74 &  \\
 & Bio\_Github Activity & 0.64 &  
                        & 2.76 \ \faArrowCircleUp & ** 
                        & 0.52 \ \faArrowCircleDown & * \\
 & Bio\_Operating Systems & 1.90 &  
                          & 2.43 \ \faArrowCircleUp & * 
                          & 1.10 &  \\
 & Bio\_Experience Years & 1.64 &  
                         & 0.22 &  
                         & 0.95 &  \\
 & Pronouns\_She/Her & 0.91 &  
                     & 0.51 \ \faArrowCircleDown & * 
                     & 0.42 \ \faArrowCircleDown & *** \\
                     
 & Country\_IN & 0.29 \ \faArrowCircleDown & ** 
              & 0.39 \ \faArrowCircleDown & * 
              & 1.09 &  \\

 & Country\_NG & 0.31 \ \faArrowCircleDown & *** 
              & 0.39 \ \faArrowCircleDown & * 
              & 0.24 \ \faArrowCircleDown & *** \\
 & Country\_UK & 0.64 &  
              & 0.83 &  
              & 1.10 &  \\

\bottomrule
\end{tabular}}
\begin{flushleft}
\footnotesize
\textit{Significance codes: 
\textasteriskcentered\textasteriskcentered\textasteriskcentered: p<0.001;\quad 
\textasteriskcentered\textasteriskcentered: p<0.01;\quad 
\textasteriskcentered: p<0.05;\quad 
.: p<0.1}
\end{flushleft}
\end{table}

Candidates from \textit{Brazil} were more likely associated to \textsl{Supportive} ($RRR = 2.55$), \textsl{Information-Seeking} ($RRR = 4.59$), \textsl{Organizational} ($RRR = 3.74$), and \textsl{Software} ($RRR = 6.48$) tasks.  

Pairwise \textit{$z$-tests} revealed several significant contrasts between pronoun and country groups in task assignments. For candidates using \textit{she/her} pronouns, assignments were significantly more likely in \textsl{Documentation} ($z = 2.88$), \textsl{General Software} ($z = 2.32$), \textsl{Information-Seeking} ($z = 2.65$), \textsl{Organizational} ($z = 2.73$), and \textsl{Other} ($z = 4.05$) compared to \textsl{Requirement-Related}. In contrast, they were less likely to be assigned to \textsl{Software} ($z = -4.75$) and \textsl{Supportive} ($z = -2.46$). Relative to \textsl{Clerical}, they showed a higher likelihood of being assigned to \textsl{Software} ($z = -3.13$), while compared to \textsl{Documentation}, \textsl{Information-Seeking}, \textsl{Organizational}, and \textsl{Supportive}, further significant differences also emerged. Finally, they were more likely to be assigned to \textsl{Other} than to \textsl{Documentation} ($z = -2.05$), \textsl{Clerical} ($z = -2.30$), and \textsl{Organizational} ($z = -2.74$).  Candidates using \textit{they/them} pronouns were more likely to be assigned to \textsl{Software} than to \textsl{Clerical} ($z = 2.14$), \textsl{Organizational} ($z = -2.48$), and \textsl{Supportive} ($z = 3.28$), and to \textsl{Documentation} rather than \textsl{Supportive} ($z = 2.30$).  
Regarding country effects, candidates from \textit{India} were more likely to be assigned to \textsl{Organizational} ($z = -2.98$), \textsl{Software} ($z = -3.52$), and \textsl{Supportive} ($z = -2.82$) compared to \textsl{Clerical}. They also showed a stronger association with \textsl{Software} relative to \textsl{Documentation} ($z = -2.32$), \textsl{Information-Seeking} ($z = -2.14$), and \textsl{Other} ($z = -3.27$), as well as to \textsl{Organizational} ($z = -2.71$) and \textsl{Supportive} ($z = -2.51$) when contrasted with \textsl{Other}. For \textit{UK} candidates, assignments were more likely in nearly all categories—\textsl{Documentation} ($z = -2.82$), \textsl{General Software} ($z = -2.80$), \textsl{Information-Seeking} ($z = -3.15$), \textsl{Organizational} ($z = -3.40$), \textsl{Other} ($z = -2.33$), \textsl{Requirement-Related} ($z = -2.46$), \textsl{Software} ($z = -3.85$), and \textsl{Supportive} ($z = -2.34$)—compared to \textsl{Clerical}. Additional contrasts showed that they were more likely to be assigned to \textsl{Software} against \textsl{Documentation} ($z = -2.00$), \textsl{Other} ($z = -2.48$), and \textsl{Supportive} ($z = 2.74$), and to \textsl{Organizational} compared to \textsl{Supportive} ($z = 2.07$).

\textbf{Claude}. The \textit{multinomial logistic regression} on Claude outputs (\autoref{tab:all_llms_multinom}) revealed significant effects for all variables. 

For \textit{confounding factors}, Claude consistently favored AI models, platforms, and communication tools, which increased assignment likelihood across categories such as \textsl{General Software}, \textsl{Requirement-Related}, \textsl{Intellectual}, \textsl{Software}, and \textsl{Other}. Collaboration tools and biography length also showed positive effects in selected categories, while biography sentiment promoted \textsl{Intellectual} tasks and company mentions promoted \textsl{Software}. 

For pronouns, candidates using \textit{she/her} were significantly less likely to be assigned to \textsl{Clerical} ($RRR = 0.07$), \textsl{Information-Seeking} ($RRR = 0.35$), \textsl{Intellectual} ($RRR = 0.12$), \textsl{Organizational} ($RRR = 0.24$), \textsl{Requirement-Related} ($RRR = 0.15$), \textsl{Software} ($RRR = 0.30$), and \textsl{Other} ($RRR = 0.42$) tasks. By contrast, they were more likely to be assigned to \textsl{General Software} tasks ($RRR = 2.22$).


The model revealed that the country also influenced task assignment. Candidates from \textit{Brazil} were more likely to be assigned to \textsl{General Software} ($RRR = 3.35$), \textsl{Organizational} ($RRR = 2.75$), \textsl{Requirement-Related} ($RRR = 3.15$), and \textsl{Software} ($RRR = 9.99$). Likewise, candidates from \textit{India} showed a higher likelihood of assignment to \textsl{Organizational} ($RRR = 3.17$) and \textsl{Requirement-Related} ($RRR = 2.67$). Those from the \textit{UK} were also more likely to be placed in \textsl{Organizational} ($RRR = 3.40$) and \textsl{Software} ($RRR = 3.57$) categories. Candidates from \textit{Nigeria} were less likely to be assigned to \textsl{Information-Seeking} ($RRR = 0.50$), \textsl{Intellectual} ($RRR = 0.28$), \textsl{Organizational} ($RRR = 0.46$), and \textsl{Other} ($RRR = 0.24$) tasks.

Pairwise \textit{$z$-tests} revealed several significant contrasts between pronoun and country categories. Candidates using \textit{she/her} pronouns were significantly more likely to be assigned to \textsl{Communicative} ($z = 2.12$), \textsl{Documentation} ($z = 2.10$), \textsl{Information-Seeking} ($z = 2.59$), and \textsl{Other} ($z = -2.98$) tasks compared to \textsl{Intellectual}. They also showed a preference for \textsl{Information-Seeking} over \textsl{Requirement-Related} ($z = 2.25$), and for \textsl{Other} over both \textsl{Requirement-Related} ($z = 2.70$) and \textsl{Version Control} ($z = 2.08$, $p = 0.038$). Candidates using \textit{he/they} pronouns were more likely to be assigned to \textsl{Clerical} rather than \textsl{Organizational} tasks ($z = 2.00$). Candidates from \textit{Brazil} were more likely to be assigned to \textsl{Software} compared to \textsl{Information-Seeking} ($z = -2.08$) and \textsl{Other} ($z = -2.68$). Those from the \textit{UK} were more likely to be placed in \textsl{Organizational} rather than \textsl{Intellectual} tasks ($z = -2.22$). For \textit{Nigeria}, multiple contrasts were significant: candidates were more likely to be assigned to \textsl{Software} compared to \textsl{Version Control} ($z = 2.02$), \textsl{Organizational} ($z = -2.07$), \textsl{Intellectual} ($z = -2.72$), \textsl{Information-Seeking} ($z = -1.97$), and \textsl{Other} ($z = -3.07$). They were also more likely to be placed in \textsl{Requirement-Related} than in \textsl{Intellectual} ($z = -2.19$) and \textsl{Other} ($z = -2.68$), in \textsl{Clerical} compared to \textsl{Intellectual} ($z = 2.15$) and \textsl{Other} ($z = 2.39$), and in \textsl{Documentation} rather than \textsl{Other} ($z = 2.22$).

\stesummarybox{\faGroup \hspace{0.05cm} RQ\textsubscript{2} -- Task Assignment.}{Task allocation was systematically shaped by pronouns, country, and profile content, with consistent trends and model-specific differences; these effects held even when accounting for confounders, showing that sensitive attributes are structurally embedded in LLM-driven allocation and raising fairness concerns. These results confirm that LLMs not only amplify demographic disparities but also reinforce stereotypes in how different groups are positioned across SE task categories.}

\section{Discussion and Implications}
Our findings uncover a complex and varied landscape of bias in LLM-driven team composition and task allocation. We stress from the outset that the following discussion should be read as a set of reflections and hypotheses informed by our statistical analyses and contextualized through recent literature, rather than as definitive causal claims. In particular, our analyses reinforce prior work~\cite{nakano2024nigerian, treude2023she} by demonstrating that biases persist even in the presence of contextual confounders. This has a general, critical implication: the \textit{decision processes adopted by LLMs do not simply reflect isolated attributes, but are shaped by the interaction of demographic and task-related variables, indicating that bias is embedded in more structural patterns of their reasoning.} In this section, we discuss the main results and draw \faHandORight \ implications for researchers and practitioners.

\smallskip
\textbf{On the interplay between pronoun and selection in a software team.}  
Across all three models, candidates using \textit{they/them} pronouns consistently faced reduced odds of selection, particularly for organizational and intellectual tasks. Similarly, \textit{she/her} profiles were less likely to be recruited and had limited access to technical roles, mirroring long-standing gendered patterns in the SE field. By contrast, candidates using more ambiguous forms such as \textit{she/they} or \textit{any/all} were sometimes favored, suggesting that LLMs may interpret inclusive or hybrid pronoun markers differently from explicit binary ones. Notably, these pronoun-related disparities persisted even when expertise was present in candidate bios, indicating that identity markers could outweigh substantive qualifications.

\steDiscussionBox{\faHandORight \ LLM designers should integrate fairness tests across diverse pronoun forms, including non-binary and hybrid usage. Practitioners should exercise caution when relying on LLM recommendations for selection, as pronoun signals may distort perceived expertise. To mitigate this risk, the adoption of double-anonymity mechanisms might help separate candidate evaluation from pronoun-related biases.}

\smallskip
\textbf{On the impact of pronoun and nationality on task allocation.}  
Bias was equally evident in task assignment. Profiles using \textit{she/her} pronouns and candidates from Nigeria were disproportionately assigned to clerical, communicative, or supportive tasks, while Brazilian, Indian, or UK candidates were more frequently placed in technical or leadership roles. This dynamic may indicate that demographic attributes condition how models interpret technical signals. For instance, Nigerian candidates with technical keywords were often selected into teams but then relegated to non-technical tasks, whereas Brazilian candidates with similar bios were channelled into technical roles. Similarly, \textit{they/them} users were consistently excluded from leadership-oriented categories, regardless of the technical expertise included in their bios. These disparities suggest that equivalent credentials may not necessarily translate into equivalent opportunities across different groups.  

\steDiscussionBox{\faHandORight \ Future research should focus on fairness-aware task allocation benchmarks that capture cross-attribute interactions. In practice, audits of LLM-based systems should adopt a multi-objective perspective, verifying not only who is selected but also whether opportunities for advancement (e.g., technical or leadership roles) are equitably distributed across groups.}

\smallskip
\textbf{Broader lessons on LLM bias.}  
Taken together, our results highlight three overarching lessons.  
First, {single-attribute analyses are insufficient}: biases become more visible when considering the joint effects of different data. Second, {LLMs risk amplifying inequities in SE contexts}: by overlooking substantive expertise and reinforcing stereotypes in task allocation, they may entrench barriers already documented in developer communities. Third, {LLM neutrality cannot be assumed}: we demonstrate that demographic markers systematically shape how technical credentials are valued.

\steDiscussionBox{\faHandORight \ For research, this requires interaction-based and multi-objective evaluation frameworks, accounting for multi-attribute interactions and socio-technical dynamics. For practice, organizations should combine LLM support with human oversight to identify when candidate expertise is undervalued due to demographic cues. For model providers, this highlights the importance of adopting mechanisms to address fairness concerns and integrating fairness benchmarks into release pipelines to mitigate real-world risks before deployment.}
\section{Threats to Validity}
\textbf{Internal Validity.}
The main threats concern, on the one hand, confounding factors that may have influenced the results. To mitigate this risk, we extracted a set of features from candidate bios using the Stack Overflow Annual Developer Survey. We incorporated them into the regression models to observe the variation. On the other hand, the variability of the LLMs' outputs. To reduce this effect, we repeated all experiments multiple times across three models. 

\noindent\textbf{External validity.}
Threats to external validity regard the generalizability. First, our dataset was derived from GitHub profiles. While GitHub is one of the most used platforms for software development, it cannot be assumed to represent the global developer population. To address this limitation, we selected the most influential countries, as identified in GitHub’s Octoverse 2024 report~\cite{github2024octoverse}, to ensure representation across diverse regions. Second, the results depend on the choice of LLMs. Clearly, this cannot cover all existing systems; therefore, we conducted the study using three different LLMs.

\noindent\textbf{Construct validity.}
These threats concern the way we defined candidate expertise and demographic attributes. Specifically, developer competence was represented through proxies such as biography length, education, and experience indicators derived from GitHub profiles, which can only approximate the effects. Ambiguities or omissions could have influenced the results. To address this limitation, we used definitions and keyword lists from validated sources.

\noindent\textbf{Conclusion validity.}
Threats to conclusion validity in our study concern the robustness of our statistical analyses and the risk of erroneous interpretation of the results. In particular, multiple comparisons increase the likelihood of Type I errors, whereas categories with limited representation may yield unstable or high-variance estimates. To address this, we employed logistic and multinomial regression models, verified model assumptions, and applied post-hoc tests with the Bonferroni correction.
\section{Conclusions}

This study examined fairness in LLM-driven team composition and task allocation, focusing on the joint effects of country and pronouns. Using three LLMs and 3,000 simulated decisions, we found systematic disparities: non-binary and female pronouns reduced selection odds, geographic patterns were inconsistent, and expertise indicators were undervalued compared to superficial keywords. These results demonstrate that bias arises not from isolated attributes but from their interplay with technical signals, underscoring the socio-technical nature of fairness challenges in LLMs.

Future work should extend this analysis to additional demographic attributes, larger and more diverse datasets, and real-world team composition scenarios. Moreover, comparative evaluations across newer LLMs and mitigation strategies are needed to design practical interventions that reduce bias in sensitive SE tasks.

\begin{acks}
We acknowledge the support of the European HORIZON-KDT-JU-2023-2-RIA research project MATISSE ``Model-based engineering of Digital Twins for early verification and validation of Industrial Systems'' (grant 101140216-2,
KDT232RIA 00017) and Project PRIN 2022 PNRR ``FRINGE: context-aware FaiRness engineerING in complex software systEms" (grant n. P2022553SL, CUP: D53D23017340001).
\end{acks}

\balance
\bibliographystyle{ACM-Reference-Format}
\bibliography{bibliography-File}

\appendix

\end{document}